# In-vivo Magnetic Resonance Imaging of GABA and Glutamate

Frederico Severo and Noam Shemesh*

*Champalimaud Research, Champalimaud Centre for the Unknown, Lisbon, Portugal*

*Corresponding author:
Noam Shemesh
Champalimaud Research, Champalimaud Centre for the Unknown, Av. Brasilia 1400-038, Lisbon, Portugal.
Phone number +351 210480000 ext #4467.
Email: Noam.Shemesh@neuro.fchampalimaud.org

**Author Contribution Statement**

NS and FS designed research; FS performed research; NS and FS contributed new tools; FS collected and analysed data; NS and FS wrote the paper.

# Abstract

Chemical Exchange Saturation Transfer (CEST) Magnetic Resonance Imaging (MRI) is a molecular imaging methodology capable of mapping brain metabolites with relatively high spatial resolution. Specificity is the main goal of such experiments; yet CEST is confounded by spectral overlap between different molecular species. Here, we overcome this major limitation using a general framework termed overlap-resolved CEST (orCEST) – a kind of spectrally-edited experiment restoring specificity. First, we present evidence revealing that CEST experiments targeting the central nervous system's primary excitatory neurotransmitter, Glutamate (GluCEST) – is significantly contaminated by gamma-aminobutyric acid (GABA) – the primary inhibitory neurotransmitter in the CNS. Then, we harness the novel orCEST methodology to separate Glutamate and – for the first time – GABA signals, thus delivering the desired specificity. In-vivo orCEST experiments resolved the rat brain's primary neurotransmitters and revealed changes in Glutamate and GABA levels upon water deprivation in thirst-related areas. orCEST's features bode well for many applications in neuroscience and biomedicine.

## Introduction

Molecular Imaging aims at spatially resolving signals originated from specific molecules with high precision and, preferably, noninvasively. Chemical Exchange Saturation Transfer (CEST) is a Magnetic Resonance Imaging (MRI) modality offering Molecular Imaging capabilities for contrasting endogenous metabolites: rather than using a genetically-labelled fluorophore or radioactive tracer, CEST MRI simply harnesses saturation (Fig. 1A) (or, less commonly, inversion or phase) transfer of signals originating from particular molecular species to the ubiquitous water signals[1,2], which can then be imaged at a high spatial resolution and completely noninvasively. For example, CEST MRI has been used to detail Glycogen variations in the liver[3]; report on glucose uptake and metabolism[4], detect myoinositol differences in grey/white matter[5]; and delineate creatine in skeletal muscle[6]. Also, CEST MRI was reported to be capable of mapping pH changes in the brain, and was used in many other applications[7–9].

Despite that specificity is the main goal of Molecular Imaging in general and CEST experiments in particular, CEST's specificity can suffer from spectral overlap of different chemical moieties (Fig. 1B). That is, two different molecules may have overlapping spectra which, upon saturation, provide partial contributions from both resonances rather than the desired one. In addition resonances from the same proton can have different offsets due to pH[10]. A striking example for spectral overlap between different molecules involves exchangeable protons of Glutamate (Glu) and Gamma-aminobutyric acid (GABA), the Central Nervous System's primary excitatory and inhibitory neurotransmitters, respectively[11]. Glutamate and GABA's roles in neural activity are paramount not only in normal brain function, but also in many brain disorders, including, among others, dementia, depression, epilepsy and schizophrenia[12–16]. Changes in Glu and GABA concentrations have been spectroscopically detected in a single voxel in disease[14], development[17], and plasticity[18], both in humans[19] and in rodents[13]. CEST MRI targeting Glutamate signals (termed GluCEST) has been recently reported[10]. However, Glutamate and GABA have very similar chemical compositions, and both exhibit CEST effects for their respective amine moieties. Contamination from other metabolites, including GABA, has been largely discounted in previous gluCEST literature[10,20–22], discarded as minimal due to (assumingly) sufficient spectral separation between the metabolite resonances; low competing CEST effect due to specific metabolite exchange rates; or even simply due to GABA's lower relative concentration in the brain[23]. Still, this ambiguity can have significant consequences in the interpretation of gluCEST data, due to the opposing neurotransmission effects. GABA CEST[24,25] has rarely been claimed compared to its Glutamate counterpart, likely due to the same overlap issue.

Here, we propose a general methodology capable of resolving spectral overlaps in CEST, termed overlap-resolved CEST (orCEST), which serves to significantly enhance CEST specificity. We investigate the extent of contamination between Glu and GABA in conventional gluCEST, and then demonstrate that, using the novel orCEST approach, both metabolites can be fully resolved and mapped separately. The first in-vivo orCEST experiments resolving Glu and GABA in-vivo are reported; changes in Glu and GABA levels, likely reflecting Glutamatergic and GABAergic mechanisms, are observed upon water-deprivation in the rat, in specific brain areas regulating thirst[26–32]. Potential implications for future studies are discussed.

# Results

## orCEST principles and optimisiation

Conventional CEST irradiation is typically performed at a particular frequency, $\omega_{peak}$ (Figure 1), aiming to maximize the contrast; however, contaminations from all the underlying signals are then also present.

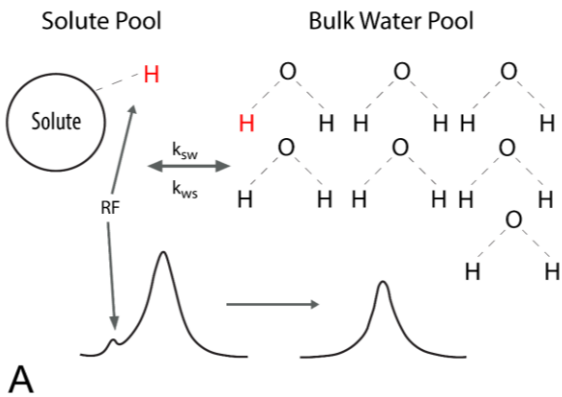
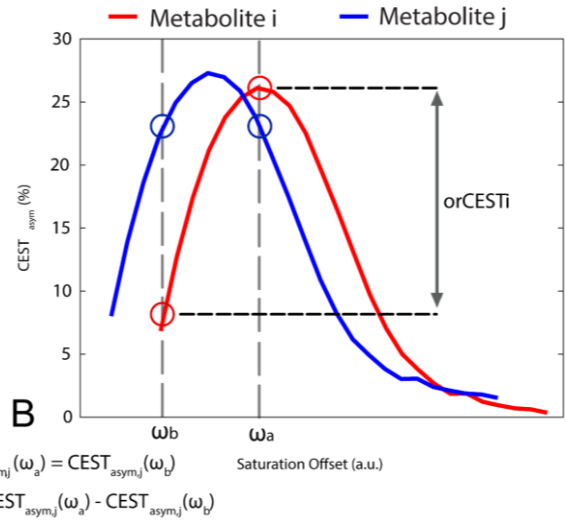
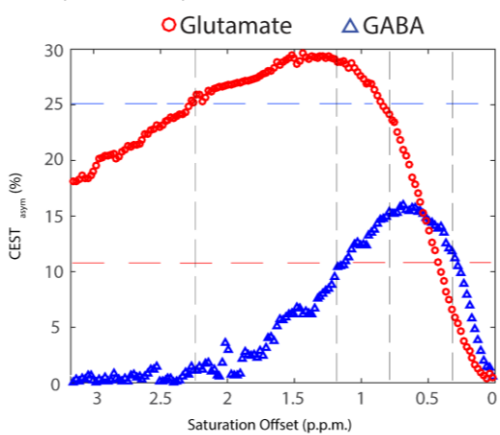
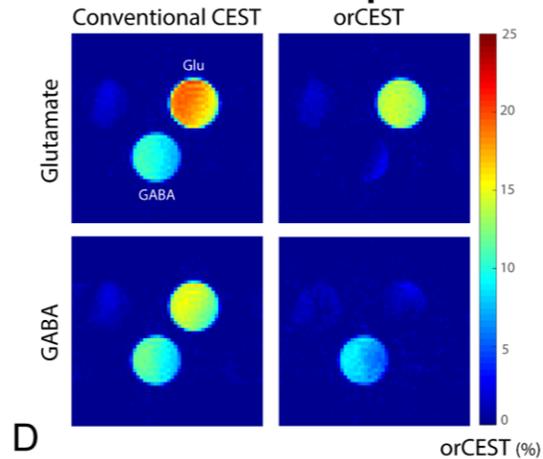

**Figure 1. CEST and orCEST mechanism of action**. **(A)** Irradiation of protons using a frequency-selective saturation pulse: Chemical exchange transfers magnetization to bulk water, resulting in indirect saturation of the bulk water that is dependent on the frequency of irradiation. **(B)** orCEST method for resolving metabolite overlap, detailing the necessary steps and conditions to increase specificity in targeting a given metabolite i by reducing metabolite's j contribution to the total CESTasym signal **(C)** CEST asymmetry curves for Glutamate and GABA at pH 7.2, with vertical grey lines depicting the acquisition frequencies necessary for orCEST calculation and blue and red horizontal lines displaying frequencies where ωa = ωb for Glutamate and GABA, respectively. **(D)** Comparison between Conventional CEST and orCEST Glutamate (10 mM) and GABA (2 mM) at pH 7.2, showing increased specificity in orCEST.

As explained above, the main goal of orCEST is to enhance the specificity of CEST experiments. The idea behind orCEST is in fact very simple, and is an analogue of spectral editing typically used in NMR[33], but applied to the CEST asymmetry (CEST$_{asym}$) spectra (Fig. 1B). If two metabolite signals partially overlap, their CEST$_{asym}$ spectra will display partially shifted peaks (Fig. 1B); if the z-spectra are a-priori known (i.e., measured in phantoms for example), symmetric points can be chosen around the peak and subtracted. Upon such subtraction, the contamination arising from the complementary molecules is effectively subtracted and cancelled out. In the orCEST experiment, maps are generated by subtracting images saturated at at $-\omega_{peak}$ and $+\omega_{peak}$, normalized to M$_0$. Thus, in orCEST, a total of four images are required, each acquired at different saturation frequencies (as well as another M$_0$ image). In the case of Glu and GABA, the resonances to be saturated were chosen as ±1.15/±0.35 p.p.m., and ±0.75/±2.15 p.p.m., for resolving Glutamate and GABA, respectively (Fig. 1c). For optimisation purposes, a range of saturation parameters were investigated, varying in number, length, power, duration, and separation of saturation pulses. From this initial screening, a subset of parameters suitable for orCEST saturation was chosen, and consisted of a train of 22 Gaussian pulses, 182 ms in duration and 15 Hz in width, and applied with 10 µT amplitudes and an inter-pulse delay of 5 ms. For other experimental details, the reader is referred to the Methods section.

**orCEST enhances specificity for Glu and GABA**

To investigate how orCEST could enhance the spectral specificity, we first show how different metabolites contribute to conventional CEST contrast (particularly, to gluCEST). Fig. 2A shows the experimental setup, incorporating different metabolites at physiological concentration (detailed in caption); Figure 2B shows the corresponding CEST$_{asym}$ curves for these metabolites, while Fig. 2C displays the ensuing (conventional) gluCEST contrast and its orCEST counterpart. Although Glutamate presents the strongest CEST signals as evident both by the intensity of the CEST$_{asym}$ curves as well as the relative amplitude in the quantified map, it is clear that other molecules cannot be ignored and that gluCEST contrast contains significant contributions from other molecules. Fig. 2D quantifies this effect, showing that around half of the signal in conventional gluCEST actually comes from other metabolites (Glu 20%, NAA 3%, GABA 7%, MI 1%, Cr 4%, Gln 7%), and, more importantly, with a considerable amount emerging from GABA – arguably the least desired contribution due to its opposing action to Glu in neurotransmission. By contrast, when the orCEST methodology is used, the specificity is significantly enhanced and the contrast reflects mainly Glutamate, containing only smaller residual signals from other molecules (Glu 16%, NAA 0.5%, GABA 2%, MI 3%, Cr 2%, Gln 1%).

Notice that other molecules are nearly in the noise level due to orCEST subtraction. Most importantly, GABA's contribution to Glutamate orCEST is now greatly reduced.

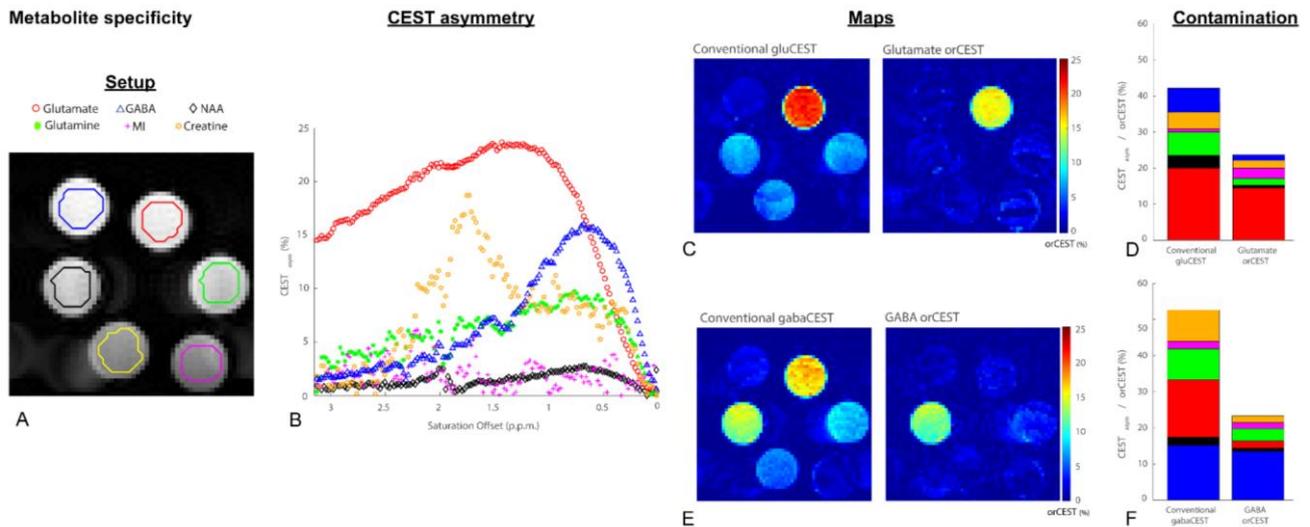

Figure 2. **Brain Metabolite contamination in CEST/orCEST. (A)** Raw data acquired for characterisation of possible metabolic contamination in orCEST. Brain metabolites at their physiological concentrations, Creatine (Cr, 4 mM), myoinositol (MI, 5 mM), N-acetyle aspartate (NAA, 8 mM), glutamine (Gln, 3 mM), Glu (10 mM) and GABA (2 mM). Image displays an acquisition without any frequency selective saturation pulse, showing no discernible difference between metabolites. **(B)** CEST asymmetry profiles of the aforementioned metabolites showing significant overlap between them. **(C)(E)** Conventional CEST and orCEST maps for Glutamate and GABA, respectively, showing improved specificity in both cases, with greatly reduced contamination from unwanted sources. **(D)(F)** Comparison between Conventional CEST and orCEST showing the massive reduction in unwanted contribution from other brain metabolites while Glutamate and GABA, respectively, maintain their CEST contrast after their respective orCEST subtraction.

As mentioned earlier, conventional gabaCEST contrast has insofar not been widely reported[24,25] as its glutamate counterpart. Fig. 2E displays the conventional gabaCEST contrast and its orCEST counterpart, showing the likely reason: conventional gabaCEST is severely contaminated with Glutamate and other molecules and in fact, the Glutamate signal is stronger than GABA's (Glu 16%, NAA 2%, GABA 15%, MI 2%, Cr 8%, Gln 9%). Indeed, GABA only contributes to about a quarter of total signal in a conventional gabaCEST acquisition (Fig. 2F), suggesting that indeed it would be difficult to ascribe specificity to gabaCEST under these conditions. When orCEST was attempted to resolve GABA, specificity was greatly enhanced in orCEST contrast. Residual contributions from Glutamate, Glutamine and Creatine are minor, with a special interest for the reduction on Glutamate contribution (Glu 2.5%, NAA 0.5%, GABA 14%, MI 1.5%, Cr 1.5%, Gln 2.5%).

It is worth noting that the subtraction inherent to orCEST incurs a small loss in sensitivity, which was found to be around 10-20%.

**Glutamate and GABA CEST - pH effects**

In the CNS, Glutamate and GABA are compartmentalized into at least two distinct main pools: extra-vesicular (mainly in the synaptic cleft and the cytosol) and intra-vesicular (within synaptic vesicles). Interestingly, each of these spaces is characterized by a very different pH: ~7.2 and ~5.5 for synaptic and intravesicular, respectively[34]. Conventional gluCEST is thought to originate mainly from pH 5.5[10]; however, given the strong effects shown above, it could be suspected that different pH environments may significantly contribute to CEST contrast.

Fig. 3A shows the setup used to test different pH effects on CEST and orCEST signals, comprising GABA and Glutamate tubes at pH 5.5 and 7.2, respectively, at physiological concentrations[23]. The corresponding CEST asymmetry curves (Fig. 3B) show that both metabolites behave very differently at different pH.

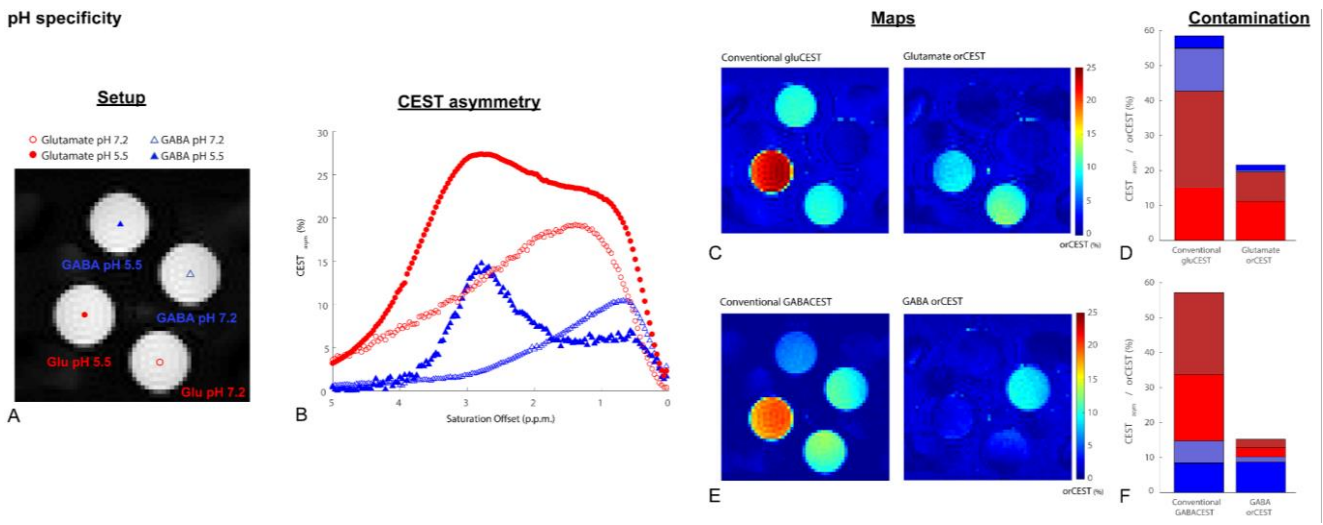

Figure 3. **Glutamate and GABA CEST/orCEST pH dependency. (A)** Raw data acquired for characterisation of orCEST, Glutamate (10 mM) and GABA (2 mM) at pH 5.5 and 7.2. In the acquisition without any frequency selective saturation pulse, no discernible difference between Glutamate and GABA can be observed at any given pH **(B)** CEST asymmetry profiles of both Glutamate and GABA, at pH 5.5 and 7.2, showing a clear overlap of the two metabolites. **(C)(E)** Conventional CEST and orCEST maps for Glutamate and GABA, respectively, showing improved specificity in both cases, with greatly reduced contamination from unwanted sources. **(D)(F)** Comparison between Conventional CEST and orCEST showing improved specificity with pH. Bright red/blue represents pH 7.2, while pale red/blue represents pH 5.5, for Glutamate and GABA, respectively.

The map corresponding to conventional gluCEST, which primarily targets Glutamate at pH 5.5 is shown in Fig. 3C, as is Glutamate orCEST. Since three tubes are clearly seen in the contrast, it becomes clear that not only does GABA contribute to gluCEST, but that glutamate at pH 7.2, which has a potentially different biological role than that of the intended contrast at pH 5.5, contributes significantly to the signal (Glu pH7.2 15%, Glu pH5.5 26%, GABA pH7.2 4%, GABA pH5.5 12%). When orCEST was applied to target Glutamate, it could clearly distinguish the population of interest: GABA tubes are not visible/quantifiable. However, it is worth noting that pH 7.2 or 5.5 both contribute to the orCEST contrast (Glu pH7.2 11%, Glu pH5.5 9%, GABA pH7.2 1%, GABA pH5.5 ~0%).

Fig. 3E shows conventional gabaCEST contrast and its orCEST counterpart, for these two pH populations. Conventional gabaCEST once again exhibits higher contrast for Glutamate than GABA itself (Glu pH7.2 17%, Glu pH5.5 23%, GABA pH7.2 8%, GABA pH5.5 5%). When GABA was targeted with orCEST, it again is able to distinguish between the populations, as evident from the lack of signal in the Glutamate tubes. Furthermore, we were able to isolate the GABA residing at pH 7.2, with very little GABA at pH 5.5 residuals (Glu pH7.2 2.5%, Glu pH5.5 2.5%, GABA pH7.2 8%, GABA pH5.5 1%), as seen in Fig. 3F.

**In vivo orCEST**. Following orCEST validation in-vitro, the first experiments in-vivo were performed in rats. Figure 4A shows the raw data from such acquisitions at three different saturation frequencies (0,-3 and +3 p.p.m.), showing the effects of direct saturation and asymmetry in saturation. In-vivo orCEST maps of Glutamate and GABA (pH 7.2) are shown in Fig. 4B and C, respectively.

Lower levels of Glutamate and GABA were found in areas rich in white matter. Glutamate levels were ~4 times higher than GABA, as expected for standard physiological levels in the brain[13,19,35,36].

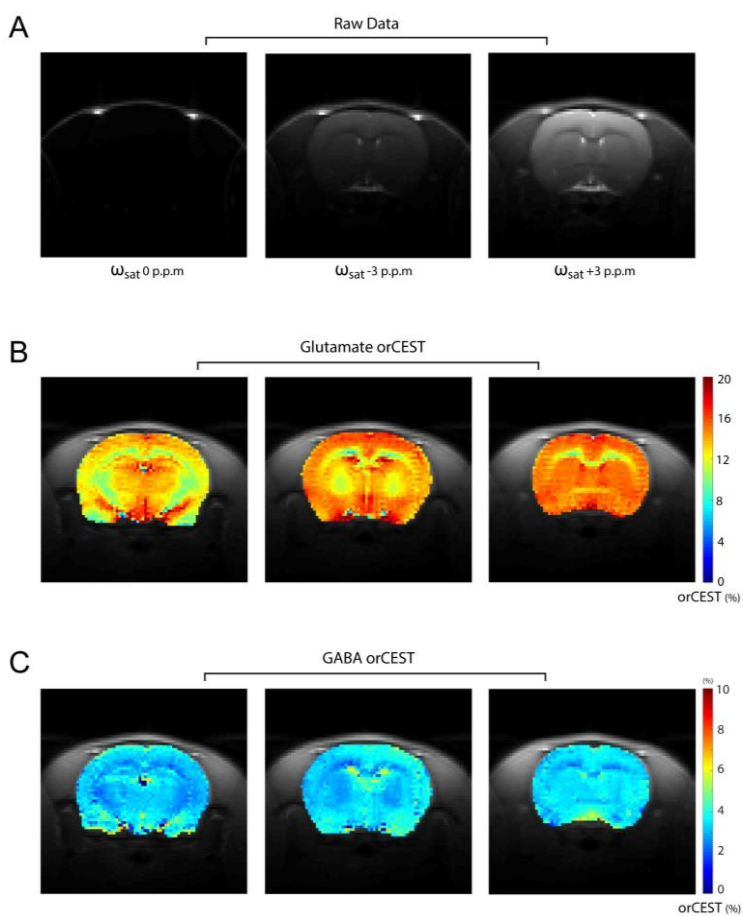

Figure 4. **Raw CEST data and orCEST maps of control rat brains** *(A)* Single animal raw data (masked) showing the effects of three different saturation frequencies (0, -3, and +3 p.p.m. from left to right), illustrating the effects of direct saturation and asymmetry in saturation effects. SNR ~25 with saturation at +3 p.p.m. **(B,C)** orCEST contrast for Glutamate and GABA in the rat brain, acquired across an axial plane at bregma 0.7/-0.9 /-1.46mm.

**Application of orCEST in water deprivation.** Thirst regulation involves several brain regions and Glutamatergic/GABAergic mechanisms. Imaging those noninvasively could make a big impact on understanding brain circuitry in-vivo. Here, orCEST was used to investigate changes in these neurotransmitters in the rat brain in vivo upon water deprivation.

A general decreasing trend is shown for both GABA and Glutamate orCEST in a ROI encompassing the whole brain (see Fig. 5A). However, specific structures, such as the Hypothalamus (Fig. 5B), Corpus Callosum (Fig. 5C) and Caudate Putamen (Fig.5D), exhibited a different behaviour upon water deprivation, with Glutamate evidencing a significant decrease for the 24H group, followed by an increase in the for the 36H group. This suggests that specific structures may have some sort of compensatory mechanism for Glutamate and GABA that is implicated in thirst regulation.

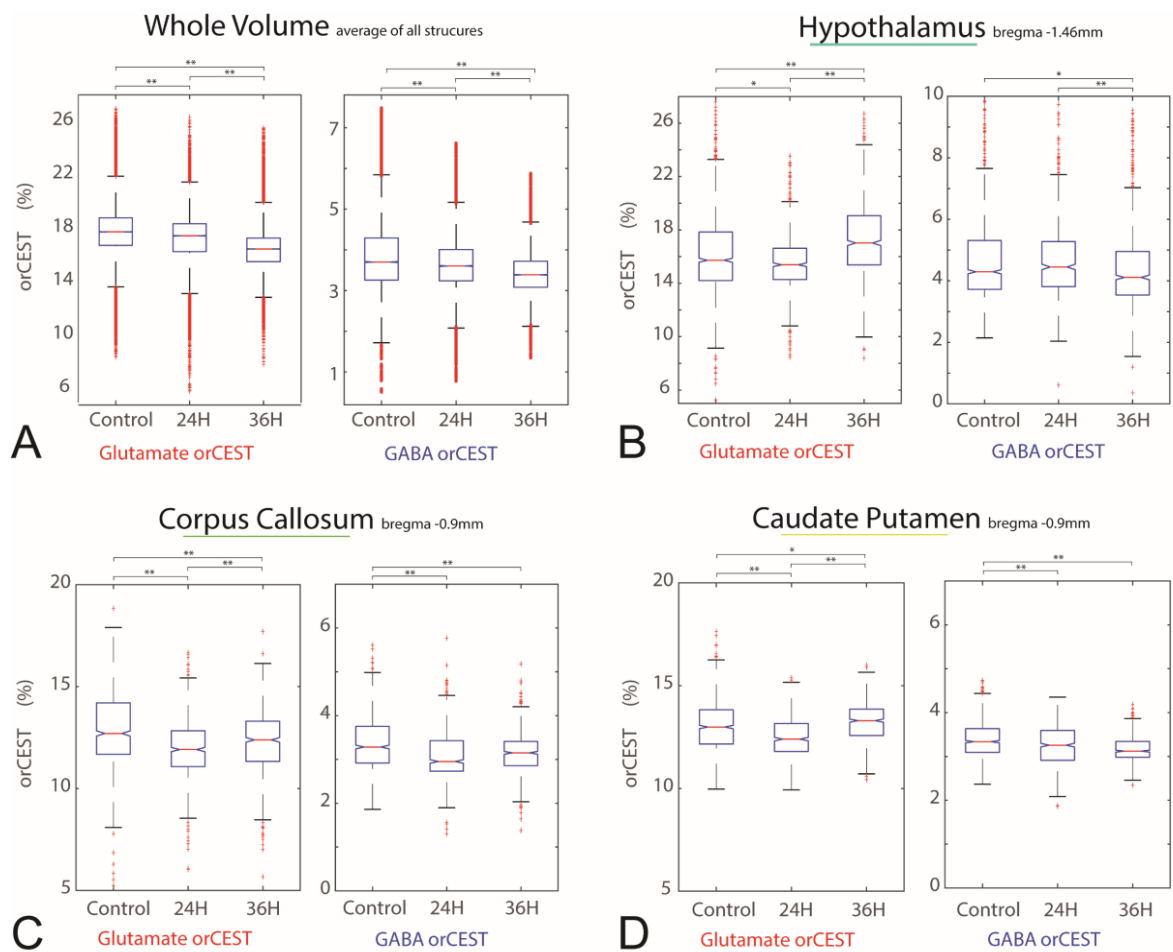

Figure 5. *orCEST variations upon water deprivation in rats at 9.4T* **(A)** Analysis on the average of all structures shows a significant downwards trend in both Glutamate and GABA between Control, 24H and 36H groups. This is the case for most brain structures. **(B,C,D)** Specific structures, such as the Hypothalamus (bregma -1,46mm), Corpus Callosum (bregma -0.9mm) and Caudate Putamen (bregma -0.9mm), exhibit a different behaviour upon water deprivation.

## Discussion

CEST experiments in general, and Glu/GABA-CEST in particular, strive to enhance the specificity of MRI experiments. However, CEST-based contrasts typically saturate at the resonance peaks to maximize contrast; in such settings, the specificity of the methodology may be significantly reduced, as here shown, due to spectral overlap from other metabolites. Many parameters can be expected to affect the degree of contamination by unwanted metabolites, including the respective frequencies and magnetic fields, as well as exchange rates, pH, and temperature. Still, it seems that, especially for Glutamate and GABA, conditions are unfavourable for conventional CEST experiments.

We presented the orCEST methodology, which aims at enhancing the specificity, even if at the expense of scan time (orCEST requires double the images, but if the resonances are judiciously chosen, orCEST will also report on double the metabolites) and some small decrease in sensitivity due to saturation off-peak. However, we have shown that orCEST provides significantly enhanced specificity which potentially resolves the molecular overlap and can partially resolve pH-related overlap, thereby providing exciting vistas for future mapping of specific (sub)cellular environments in-vivo.

Our findings are important in the context of interpretation of CEST contrasts in general, and GluCEST in particular. Indeed, GluCEST contrasts in different conditions, such as stroke models, epilepsy, psychosis or Huntington's disease[10,21,22,37] have shown markedly different contrasts. Given the GABA contributions to GluCEST here shown, one must consider whether GABAergic contributions could have played an important role in these contrasts.

As a general method for measuring long term Glutamate/GABA changes in the brain, orCEST may prove to be quite a valuable tool due to its increased specificity, especially for GABA measurements in vivo. GABA also plays a very important role in disease[38–40], and studying its variation may become a possible with orCEST. Epileptic drugs for example, often rely on the inhibition of release/reuptake of GABA in gabaergic neurons[41], and less so on Glutamatergic processes. orCEST could provide specificity and thereby facilitate direct imaging of such effects. Indeed, in our water deprivation model, with the exception of three major structures, Hypothalamus (bregma -1,46mm), Corpus Callosum (bregma -0.9mm), and Caudate Putamen (bregma -0.9mm) (supplementary Fig. 1), most brain regions show a consistent decrease in both Glutamate and GABA orCEST. Previous studies on rat models have reported an increase in hypothalamic Glutamate, upon acute dehydration over a 48h period[42], and for both Glutamate and GABA upon 7 to 10 days of chronic dehydration (2% NaCl in water)[43], evidenced by an increase in the frequency of spontaneous excitatory and inhibitory postsynaptic currents that can perhaps be attributed to elevated levels of both neurotransmitters. Because these experiments do not include data on short-term water deprivation (24 hours), no comparison can be made on the Glutamate/GABA decrease seen in our data for the first 24 hours post water deprivation. A similar variation for Glutamate and GABA was found in our data for the Caudate Putamen, which has mostly been associated with the initiation and control of movement related to water consumption, rather than more specifically to the control of thirst[44].

There are, of course, drawbacks associated with orCEST. First, twice the number of images need to be acquired, which can present practical limitations. Secondly, orCEST (and the majority of the CEST applications for that matter) relies on a stable pH, which should be taken into account in models where pH changes are expected. Changes in pH are responsible for the phenomenon of intermediate-to-fast-exchange ($k \geq \Delta\omega$)-mediated chemical shift averaging[2], which ultimately results in a frequency shift in the CEST asymmetry curve. While this effect has been taken advantage of in the past[8,9,45–47], considering the reliance of orCEST on the subtraction of specific frequencies, it is certainly important to consider it in order to choose a suitable model to investigate. Other factors, such as temperature and concentration can also affect the exchangeable resonances, albeit to smaller consequence. Finally, just as in CEST, asymmetric magnetization transfer (MT) effects may contaminate orCEST signals because of potential magnetization exchange between water molecules bound to larger macromolecules in solid or semisolid phases and free water.

In conclusion, orCEST significantly enhances the specificity in CEST experiments. Our application for Glu and GABA opens an array of possibilities for studies on the inhibitory gabaergic system, while also increasing the specificity for Glutamate imaging using MRI, in a non-invasive manner. These augur well for future molecular imaging studies.

## Materials and Methods

Phantoms

Metabolite solutions were prepared using 1xPBS (pH ~ 7.2) as the solvent and their pH was controlled by titration with 0.1 M HCl or 0.1 M NaOH. The pH was measured using a Sartorius docu-pH Meter [Sartorius AG, Goettingen, Germany] accurate to 0.1 pH units.

These samples were added to small test tubes (3 mm diameter), and immersed inside a larger (10 mm) NMR tube filled with Fluorinert [Sigma-Aldrich Chemie Gmbh, Munich, Germany], to match susceptibility. To assess pH dependency and variation, two Glutamate solutions with pH 7.2 and pH 5.5, and two GABA solutions with pH 7.2 and pH 5.5, were prepared. The pH values of 7.2 and 5.5 were chosen in accordance to the intra and extra vesicular environment in which these metabolites can be found[34]. To evaluate the contribution of other major brain metabolites in orCEST signals, solutions of the following metabolites in their typical range of physiological concentrations[23] were prepared: creatine (Cr, 4 mM), myoinositol (MI, 5 mM), N-acetyle aspartate (NAA, 8 mM), glutamine (Gln, 3 mM), Glu (10 mM) and GABA (2 mM). All samples were adjusted to pH 7.2 just prior to imaging.

In vitro experiments were performed on a Bruker Ascend Aeon 16.4 T vertical scanner (Bruker, Karlsruhe, Germany), on a micro5 probe equipped with a 10 mm coil, and a gradient system capable of producing 3000 mT/m in all directions, or, for the metabolic contamination phantom experiment, on a micro2.5 probe equipped with a 25 mm coil and a gradient system capable of producing up to 1500mT/m in all directions.

Phantom experiments were performed at 37°C, using a modified Half-Fourier Acquisition Single-shot Turbo spin Echo imaging (HASTE)[48] sequence preceded by a CEST module. Imaging parameters were as follows, TR/TE = 15000 / 15 ms, FOV = 9x9 mm, slice thickness = 3 mm, Matrix size = 56x56, number of π-pulses = 33, partial Fourier factor = 1.8, number of averages = 2, with a total scan time per run of 30 s. All experiments were acquired in a fully relaxed state (TR > 5T1). For optimisation purposes, a range of saturation parameters were investigated, varying in number, length, power, duration, and separation of saturation pulses. From this initial screening, a subset of parameters suitable for orCEST imaging were chosen, and consisted of a train of 22 Gaussian pulses, 182 ms in duration and 15 Hz in width, and applied with 10 µT amplitudes and an inter-pulse delay of 5 ms. Raw CEST acquisition in phantoms consisted of an array of varying saturation offset frequencies ranging from -3.25 to +3.25 p.p.m. (-5 to +5 p.p.m. to assess pH dependency), varied in steps of 0.02 p.p.m. (15Hz at 16.4T). An $M_0$ image (identical sequence but with zero power on the saturation pulses) for normalization was also acquired.

**In Vivo**

Long-Evans male rats (6-9 weeks old and weighing 220-320g), housed 2 per cage and allowed *ad libitum* access to food and water before the study started were randomly split into three groups, Control (n=16), 24H Water Deprived (24H, n=16), and 36H Water Deprived (36H, n=16). The first group was allowed ad libitum access to food and water, while the second and third were water deprived for 24h and 36h, respectively. In the beginning of water deprivation, healthy animals were weighted and the water bottles on the cages were made unavailable. All animals were monitored every 12 hours and the continuation of water restriction was decided according to welfare and health of individuals.

Animals were anesthetized with 5% isofluorane for induction, and ~2.5% for maintenance delivered through a nosecone and carried by 95% Oxygen (medical air). Once anesthetized, the animals were placed in a dedicated animal bed and inserted vertically to the MRI scanner. Breathing rate and temperature were monitored throughout the experiment using a SA Instruments Model 1030 Monitoring & Gating system (SA Instruments Inc, NY, USA) interfaced with the scanner.

In vivo experiments were conducted on a Bruker BioSpec 9.4T equipped with a Rat cryoprobe and a gradient system capable of producing up to 660 mT/m in all directions. All acquisitions were performed with the animal kept at a stable temperature of 37°C, using a modified Half-Fourier Acquisition Single-shot Turbo spin Echo imaging (HASTE)[48] sequence preceded by a CEST module. Imaging parameters were: TR/TE = 10000/14.7ms, FOV = 24x24mm, slice thickness = 1.25mm, Matrix size = 80x80, number of refocusing pulses=44, partial Fourier factor=1.8, number of averages = 10, with a total scan time per run of 1m40s. All experiments were acquired in a fully relaxed state (TR>5T1). Saturation consisted of 22 Gaussian pulses, 182ms/15Hz, 10µT, with an interpulse delay of 5ms. A total of three slices were acquired across an axial plane at bregma 0.7/-0.9 /-1.46mm. B0 and B1 profiles were optimized per slice prior to each acquisition. Besides the CEST data acquired at multiple , data for B1 and B0 correction was also acquired. One-way ANOVA corrected for multiple comparisons was used to analyse the acquired data.

**orCEST Quantification**

CEST contrast is determined through the collection of a Z-spectrum, a plot of the normalized z magnetization of water, as described by

[Eq.1]
$$M_{sat(\omega)}/M_0$$

where $M_{sat}(\omega)$ and $M_0$ represents magnetization saturated at frequency ω and unsaturated magnetization, respectively.

For conventional CEST applications, each metabolite can be characterised by its own CEST asymmetry spectrum,

[Eq.2]
$$CEST_{asym,i} = \frac{M_{sat}(-\omega) - M_{sat}(\omega)}{M_0} \times 100$$

where *i* represents the *i*th metabolite. Such $CEST_{asym}$ spectra tend to exhibit local maxima; if the peak arises from a single metabolite, its saturation at $\omega_{peak,i}$ should provide specificity to metabolite *i*. However, if another metabolite *j* has overlapping $CEST_{asym}$ signals, saturation at $\omega_{peak,i}$ would inherently lead to contamination from metabolite *j*.

One potential solution for this contamination, is to use a kind of "spectral editing": that is, find a way to subtract out one of the metabolite signals. Due to the peaked shape of the spectra, there should exist at least two frequencies for which $CEST_{asym}$ intensities are identical

[Eq. 3]
$$CEST_{asym,i}(\omega_a) = CEST_{asym,i}(\omega_b)$$

where $\omega_a \neq \omega_b$ (Fig. 1B). Similarly, for metabolite $j \neq i$ with a partially overlapping peaked $CEST_{asym}$ spectrum, one can find frequencies $\omega_{c,d}$ such that $\omega_c \neq \omega_d$ ($\neq \omega_a \neq \omega_b$) for which

[Eq. 4]
$$CEST_{asym,i}(\omega_c) = CEST_{asym,i}(\omega_d)$$

Each frequency pair can thus serve as subtraction points, for which signals from one specific metabolite will be nulled; that is, one can null the effects of metabolite *i* while preserving contrast from metabolite *j* (Fig. 1B). The task for such overlap-resolved CEST (orCEST) spectra is then to select frequency pairs such that contribution from unwanted sources is minimized, while preserving the maximum possible CEST effect for the desired metabolite. For peak signals *i,j* and nulling frequencies $\omega_c$ and $\omega_d$ nulling signal *j* and $\omega_a$ and $\omega_b$ nulling signal *i* (Fig. 1B):

[Eq. 5]
$$orCEST_i = CEST_{asym}(\omega_c) - CEST_{asym}(\omega_d)$$
$$orCEST_j = CEST_{asym}(\omega_a) - CEST_{asym}(\omega_b)$$

Metabolites *i* and *j* have thus been resolved with considerably more specificity.

**Data Analysis**

Image processing and data analysis were done in MATLAB r2015b® (The Mathworks, Nattick, MA, USA), using custom-written code. Intermediate frequency CEST-weighted images were interpolated with spline fitting to obtain the signal intensity S at the desired offsets $\omega_{desired}$,

[Eq.6]
$$S_{interpolated} = spline[\omega_{exp}, S_{exp}(x, y, \omega_{exp}), \omega_{desired}(x, y)]$$

To correct for spatial $B_0$ field variations, WASSR correction[49] was used. The irradiated frequencies are adjusted by

[Eq.7]
$$\omega_{corrected}(x, y) = \omega_{exp}(x, y) - \delta\omega_0(x, y)$$

All in vivo data were denoised using Veraart's method of Marchenko-Pastur distribution removal in PCA of redundant datasets[50], and then corrected for Gibbs ringing using the Gibbs unringing algorithm presented by Kellner et al.[51], prior to further analysis. ROIs for in vivo orCEST data were drawn manually with the help of a rat brain atlas[52], and data was tested for normality using One-Sample Kolmogorov-Smirnov test. Grubbs' test for outliers was applied and new points were interpolated if needed. Equivalent ROIs from different animals within the same group were clustered and a one-way ANOVA was applied on both Control, 24H and 36H groups, with a Post Hoc multiple comparisons test (Bonferroni correction). Plots were generated from this analysis pipeline.

# References

1. Liu, G., Song, X., Chan, K. W. Y. & Mcmahon, M. T. Nuts and bolts of chemical exchange saturation transfer MRI. *NMR Biomed.* **26**, 810–828 (2013).
2. Van Zijl, P. C. M. & Yadav, N. N. Chemical exchange saturation transfer (CEST): What is in a name and what isn't? *Magn. Reson. Med.* **65**, 927–948 (2011).
3. Shah, T. *et al.* CEST-FISP: A novel technique for rapid chemical exchange saturation transfer MRI at 7 T. *Magn. Reson. Med.* **65**, 432–437 (2011).
4. Nasrallah, F. A., Pages, G., Kuchel, P. W., Golay, X. & Chuang, K. H. Imaging brain deoxyglucose uptake and metabolism by glucoCEST MRI. *J Cereb Blood Flow Metab* **33**, 1270–1278 (2013).
5. Haris, M. *et al.* MICEST: A potential tool for non-invasive detection of molecular changes in Alzheimer's disease. *J. Neurosci. Methods* **212**, 87–93 (2013).
6. Kogan, F. *et al.* In vivo chemical exchange saturation transfer imaging of creatine (CrCEST) in skeletal muscle at 3T. *J. Magn. Reson. Imaging* **40**, 596–602 (2014).
7. Chen, L. Q. & Pagel, M. D. Evaluating pH in the Extracellular Tumor Microenvironment Using CEST MRI and Other Imaging Methods. *Adv. Radiol.* **2015**, 1–25 (2015).
8. Sun, P. Z. & Gregory Sorensen, A. Imaging pH using the chemical exchange saturation transfer (CEST) MRI: Correction of concomitant RF irradiation effects to quantify cest MRI for chemical exchange rate and pH. *Magn. Reson. Med.* **60**, 390–397 (2008).
9. Wermter, F. C., Bock, C. & Dreher, W. Investigating GluCEST and its specificity for pH mapping at low temperatures. *NMR Biomed.* **28**, 1507–1517 (2015).
10. Cai, K. *et al.* Magnetic resonance imaging of glutamate. *Nat. Med.* **18**, 302–306 (2012).
11. Kandel, E. R., Schwartz, J. H. & Jessell, T. M. *Principles of Neural Science*. *Neurology* **3**, (2000).
12. Bradford, H. F. Glutamate, GABA and epilepsy. *Progress in Neurobiology* **47**, 477–511 (1995).
13. Jo, S. *et al.* GABA from reactive astrocytes impairs memory in mouse models of Alzheimer's disease. *Nat. Med.* **20**, 886–96 (2014).
14. Błaszczyk, J. W. Parkinson's disease and neurodegeneration: GABA-collapse hypothesis. *Front. Neurosci.* **10**, (2016).
15. Greenberg, D. A. Glutamate and Parkinson's disease. *Annals of Neurology* **35**, 639–639 (1994).
16. Walton, H. S. & Dodd, P. R. Glutamate-glutamine cycling in Alzheimer's disease. *Neurochem. Int.* **50**, 1052–1066 (2007).
17. Ben-Ari, Y. Excitatory actions of gaba during development: the nature of the nurture. *Nat Rev Neurosci* **3**, 728–739 (2002).
18. Stagg, C. J. Magnetic Resonance Spectroscopy as a tool to study the role of GABA in motor-cortical plasticity. *NeuroImage* **86**, 19–27 (2014).
19. Houtepen, L. C. *et al.* Acute stress effects on GABA and glutamate levels in the prefrontal cortex: A 7T 1H-magnetic resonance spectroscopy study. *NeuroImage Clin.* **14**, 195–200 (2017).
20. Bagga, P. *et al.* Mapping the alterations in glutamate with GluCEST MRI in a mouse model of dopamine deficiency. *J. Neurochem.* **139**, 432–439 (2016).
21. Roalf, D. R. *et al.* Glutamate imaging (GluCEST) reveals lower brain GluCEST contrast in patients on the psychosis spectrum. *Mol. Psychiatry* 1–8 (2017). doi:10.1038/mp.2016.258
22. Davis, K. A. *et al.* Glutamate imaging (GluCEST) lateralizes epileptic foci in nonlesional temporal lobe epilepsy. *Sci. Transl. Med.* **7**, 309ra161 (2015).
23. Duarte, J. M. N., Lei, H., Mlynárik, V. & Gruetter, R. The neurochemical profile quantified by in vivo 1H NMR spectroscopy. *NeuroImage* **61**, 342–362 (2012).
24. Lee, D. H. *et al.* Changes to gamma-aminobutyric acid levels during short-term epileptiform activity in a kainic acid-induced rat model of status epilepticus: A chemical exchange saturation transfer imaging study. *Brain Res.* (2019). doi:10.1016/j.brainres.2019.04.010
25. Cai, K., Haris, M., Singh, A., Kogan, F. & Waghray, P. Magnetic Resonance Imaging of the Neurotransmitter GABA in-vivo. *Proc. Intl. Soc. Mag. Reson. Med.* **18**, 2983 (2010).
26. Abbott, S. B. G., Machado, N. L. S., Geerling, J. C. & Saper, C. B. Reciprocal Control of Drinking Behavior by Median Preoptic Neurons in Mice. *J. Neurosci.* **36**, 8228–8237 (2016).
27. Coburn, P. C. & Striker, E. M. Osmoregulatory Thirst in Rats After Lateral Preoptic Lesions. **92**, 350–361 (1978).
28. Oka, Y., Ye, M. & Zuker, C. S. Thirst driving and suppressing signals encoded by distinct neural populations in the brain. *Nature* **520**, 349–352 (2015).
29. Cambiasso, M. J. The Involvement of the Hypothalamic Preoptic Area on the Regulation of Thirst in the Rat. **195**,

190–195 (1992).
30. Zimmerman, C. A., Leib, D. E. & Knight, Z. A. Neural circuits underlying thirst and fluid homeostasis. *Nat. Rev. Neurosci.* **18**, 459–469 (2017).
31. Leshem, M. & Epstein, A. N. Thirst-induced anorexias and the ontogeny of thirst in the rat. *Dev. Psychobiol.* **21**, 651–662 (1988).
32. Sharifkhodaei, Z. Interaction between Dopaminergic and Angiotensinergic Systems on Thirst in Adult Male Rats. *Neurosci. Med.* **03**, 75–82 (2012).
33. Bogner, W., Hangel, G., Esmaeili, M. & Andronesi, O. C. 1D-spectral editing and 2D multispectral in vivo 1H-MRS and 1H-MRSI - Methods and applications. *Anal. Biochem.* **529**, 48–64 (2017).
34. Atluri, P. P. & Ryan, T. a. The kinetics of synaptic vesicle reacidification at hippocampal nerve terminals. *J. Neurosci.* **26**, 2313–20 (2006).
35. Petroff, O. A. C. GABA and glutamate in the human brain. *Neuroscientist* (2002). doi:10.1177/1073858402238515
36. Agarwal, N. & Renshaw, P. F. Proton MR spectroscopy - Detectable major neurotransmitters of the brain: Biology and possible clinical applications. *American Journal of Neuroradiology* **33**, 595–602 (2012).
37. Pépin, J. *et al.* In vivo imaging of brain glutamate defects in a knock-in mouse model of Huntington's disease. *Neuroimage* **139**, 53–64 (2016).
38. Reynolds, G. & Sally, P. Brain GABA Levels in Asymptomatic Huntington's Disease. *N. Engl. J. Med.* **323**, 682–682 (1990).
39. Blicher, J. U. *et al.* GABA levels are decreased after stroke and GABA changes during rehabilitation correlate with motor improvement. *Neurorehabil.Neural Repair.* **29**, 278–286 (2015).
40. Orhan, F. *et al.* CSF GABA is reduced in first-episode psychosis and associates to symptom severity. *Mol. Psychiatry* 1–7 (2017). doi:10.1038/mp.2017.25
41. Treiman, D. M. GABAergic mechanisms in epilepsy. in *Epilepsia* **42**, 8–12 (2001).
42. Holbein WH, Bardgett ME, Toney, G. Blood Pressure is maintained during dehydration by hypothalamic paraventricular nucleus-driven tonic sympathetic nerve activity. *J Physiol* **592(Pt 17)**, 3783–3799 (2014).
43. Di, S. & Tasker, J. G. Dehydration-induced synaptic plasticity in magnocellular neurons of the hypothalamic supraoptic nucleus. *Endocrinology* (2004). doi:10.1210/en.2004-0702
44. Rolls BJ, W. R. Role of angiotensin in thirst. *Pharmacol Biochem Behav* **Mar;6(3):2**, (1977).
45. Harris, R. J. *et al.* PH-weighted molecular imaging of gliomas using amine chemical exchange saturation transfer MRI. *Neuro. Oncol.* **17**, 1514–1524 (2015).
46. McVicar, N. *et al.* Quantitative tissue pH measurement during cerebral ischemia using amine and amide concentration-independent detection (AACID) with MRI. *J. Cereb. Blood Flow Metab.* **34**, 690–8 (2014).
47. Zhou, J., Payen, J.-F., Wilson, D., Traystman, R. & van Zijl, P. C. M. Using the amide proton signals of intracellular proteins and peptides to detect pH effects in MRI. *Nat. Med.* **9**, 1085–1090 (2003).
48. Patel, M. R., Klufas, R. A., Alberico, R. A. & Edelman, R. R. Half-fourier acquisition single-shot turbo spin-echo (HASTE) MR: Comparison with fast spin-echo MR in diseases of the brain. *Am. J. Neuroradiol.* **18**, 1635–1640 (1997).
49. Kim, M., Gillen, J., Landman, B. A., Zhou, J. & Van Zijl, P. C. M. Water saturation shift referencing (WASSR) for chemical exchange saturation transfer (CEST) experiments. *Magn. Reson. Med.* **61**, 1441–1450 (2009).
50. Veraart, J. *et al.* Denoising of diffusion MRI using random matrix theory. *Neuroimage* **142**, 394–406 (2016).
51. Kellner, E., Dhital, B., Kiselev, V. G. & Reisert, M. Gibbs-ringing artifact removal based on local subvoxel-shifts. *Magn. Reson. Med.* **76**, 1574–1581 (2016).
52. Paxinos, G. & Watson, C. The rat brain in stereotaxic coordinates (6th ed.). *Acad. Press* (2007).
53. Zu, Z. *et al.* Multi-angle ratiometric approach to measure chemical exchange in amide proton transfer imaging. *Magn. Reson. Med.* **68**, 711–719 (2012).
54. Mori, S., Johnson, M., Berg, J. M. & van Zijl, P. C. M. Water Exchange Filter (WEX Filter) for Nuclear Magnetic Resonance Studies of Macromolecules. *J. Am. Chem. Soc.* **116**, 11982–11984 (1994).
55. Chan, K. W. Y. *et al.* CEST-MRI detects metabolite levels altered by breast cancer cell aggressiveness and chemotherapy response. *NMR Biomed.* **29**, 806–816 (2016).

# Figure Captions

Fig. 1 *CEST and orCEST mechanism of action* **(A)** Irradiation of protons using a frequency-selective saturation pulse: Chemical exchange transfers magnetization to bulk water, resulting in indirect saturation of the bulk water that is dependent on the frequency of irradiation. **(B)** orCEST method for resolving metabolite overlap, detailing the necessary steps and conditions to increase specificity in targeting a given metabolite i by reducing metabolite's j contribution to the total $CEST_{asym}$ signal **(C)** CEST asymmetry curves for Glutamate and GABA at pH 7.2, with vertical grey lines depicting the acquisition frequencies necessary for orCEST calculation and blue and red horizontal lines displaying frequencies where $\omega_a = \omega_b$ for Glutamate and GABA, respectively. **(D)** Comparison between Conventional CEST and orCEST Glutamate (10 mM) and GABA (2 mM) at pH 7.2, showing increased specificity in orCEST.

Fig. 2 **Brain *Metabolite contamination in CEST/orCEST at 16.4T*** **(A)** Raw data acquired for characterisation of possible metabolic contamination in orCEST. Brain metabolites at their physiological concentrations, Creatine (Cr, 4 mM), myoinositol (MI, 5 mM), N-acetyle aspartate (NAA, 8 mM), glutamine (Gln, 3 mM), Glu (10 mM) and GABA (2 mM). Image displays an acquisition without any frequency selective saturation pulse, showing no discernible difference between metabolites. **(B)** CEST asymmetry profiles of the aforementioned metabolites showing significant overlap between them. **(C)(E)** Conventional CEST and orCEST maps for Glutamate and GABA, respectively, showing improved specificity in both cases, with greatly reduced contamination from unwanted sources. **(D)(F)** Comparison between Conventional CEST and orCEST showing the massive reduction in unwanted contribution from other brain metabolites while Glutamate and GABA, respectively, maintain their CEST contrast after their respective orCEST subtraction.

Fig. 3 *Glutamate and GABA CEST/orCEST pH dependency at 16.4T* **(A)** Raw data acquired for characterisation of orCEST, Glutamate (10 mM) and GABA (2 mM) at pH 5.5 and 7.2. In the acquisition without any frequency selective saturation pulse, no discernible difference between Glutamate and GABA can be observed at any given pH **(B)** CEST asymmetry profiles of both Glutamate and GABA, at pH 5.5 and 7.2, showing a clear overlap of the two metabolites. **(C)(E)** Conventional CEST and orCEST maps for Glutamate and GABA, respectively, showing improved specificity in both cases, with greatly reduced contamination from unwanted sources. **(D)(F)** Comparison between Conventional CEST and orCEST showing improved specificity with pH. Bright red/blue represents pH 7.2, while pale red/blue represents pH 5.5, for Glutamate and GABA, respectively.

Fig 4. *Raw CEST data and orCEST maps of control rat brains at 9.4T* **(A)** Single animal raw data (masked) showing the effects of three different saturation frequencies (0, -3, and +3 p.p.m. from left to right), illustrating the effects of direct saturation and asymmetry in saturation effects. SNR ~25 with saturation at +3 p.p.m. **(B,C)** orCEST contrast for Glutamate and GABA in the rat brain, acquired across an axial plane at bregma 0.7/-0.9 /-1.46mm.

Fig 5. *orCEST variations upon water deprivation in rats at 9.4T* **(A)** Analysis on the average of all structures shows a significant downwards trend in both Glutamate and GABA between Control, 24H and 36H groups. This is the case for most brain structures. **(B,C,D)** Specific structures, such as the Hypothalamus (bregma -1,46mm), Corpus Callosum (bregma -0.9mm) and Caudate Putamen (bregma -0.9mm), exhibit a different behaviour upon water deprivation.